\documentclass[a4paper,11pt]{article}
\pdfoutput=1 

\usepackage{jinstpub} 

\usepackage[normalem]{ulem}
\usepackage{amsmath}
\usepackage{amssymb}
\usepackage{epsfig}
\usepackage{graphicx}
\usepackage{tabularx}
\usepackage{colortbl}
\usepackage{xcolor}

\usepackage{comment}
\graphicspath{{figs/}}

\title{Charge collection efficiency in back-illuminated Charge-Coupled Devices.}

\author[a]{Guillermo Fernandez-Moroni}
\author[a,b]{, Kevin Andersson}
\author[b]{, Ana Botti}
\author[a]{, Juan Estrada}
\author[a,b]{, Dario Rodrigues}
\author[a]{and Javier Tiffenberg.}
\affiliation[a]{\normalsize\it  Fermi National Accelerator Laboratory, PO Box 500, Batavia IL, 60510}
\affiliation[b]{\normalsize\it Department of Physics, FCEN, University of Buenos Aires and IFIBA, CONICET, Buenos Aires, Argentina}

\emailAdd{gfmoroni@fnal.gov}

\abstract{Low noise CCDs fully-depleted up to 675 micrometers have been identified as a unique tool for Dark Matter searches and low energy neutrino physics. The charge collection efficiency (CCE) for these detectors is a critical parameter for the performance of future experiments. We present here a new technique to characterize CCE in back-illuminated CCDs based on soft X-rays. This technique is used to characterize two different detector designs. The results demonstrate the importance of the backside processing for detection near threshold, showing that a recombination layer of a few microns significantly distorts the low energy spectrum. The studies demonstrate that the region of partial charge collection can be reduced to less than 1 $\mu$m thickness with adequate backside processing.}


\proceeding{FERMILAB-PUB-20-278-E}
  
\begin{document}
\maketitle
\flushbottom

\section{Thick Fully-Depleted CCDs for Dark Matter and neutrino experiments}


Charged Coupled Devices (CCD) with low readout noise and large active volume have been identified among the most promising detector technologies for the low mass direct dark matter search experiments, probing electron and nuclear recoils from sub-GeV DM \cite{Aguilar-Arevalo:2016ndq, Aguilar-Arevalo:2016zop,Aguilar-Arevalo:2019wdi, Crisler:2018gci,Abramoff:2019dfb}. The recent development of the Skipper-CCD \cite{Tiffenberg:2017aac,Sensei2020}  demonstrated the ability to measure ionization events with sub-electron noise extending the reach of this technology to unprecedented low energies. Experiments based on this technology are planned for the coming years with total CCD active mass going from 100 grams to several kilograms \cite{DAMIC-M,Oscura}. At the same time the low noise CCD technology has been implemented in low energy neutrino experiments \cite{CONNIE_2019,CONNIE_2020} and are planned for future developments\cite{Violeta}.


There are several key performance parameters for the CCD sensors in future developments that are part of a significant R\&D effort for future projects \cite{DAMIC-M,Oscura,Violeta}. The most important performance requirements are the pixel dark current \cite{Sensei2020}, readout noise optimization \cite{LTA}, Fano factor \cite{Rodrigues2020}  and charge transport in the sensor \cite{SofoDiffusion}. 

The Charge Collection Efficiency (CCE) is defined as the fraction of the total charge produced during a ionization event that is collected in the CCD pixel for later readout. For a fully depleted detector, with a large electric field, CCE is approximately 100\% \cite{janesick2001scientific} for the full active volume. 
In regions of the detector with lower electric field, CCE could be less than 100\% due to charge recombination. Regions of partial CCE distort the measured spectrum of ionization events, affecting energy calibration and particle identification.

Back illuminated CCDs in astronomy are treated to have a thin entrance window for light, with low reflectivity. This is specially important when detectors are used for wavelength shorter than 500 nm  \cite{JPLdelta_1994,JPLdelta_2016,DESICCD2017}. A 500 nm photon has an absorption length of 1 $\mu$m in Silicon, and any layer with partial charge collection (PCC) on the back surface will degrade the detection efficiency for blue light. The measurements presented in Ref.\cite{LBNLQE} compare the detection efficiency for visible photons with the reflectivity. These studies show that all photons with wavelength longer than 500 nm are fully detected, unless they are reflected on the back surface. These results show that the bulk of the detector has 100\% CCE, and that any recombination on these sensors occurs only on the first 1$\mu$m near the back surface.

For thick CCDs, as those used in dark matter \cite{Aguilar-Arevalo:2016ndq, Aguilar-Arevalo:2016zop,Aguilar-Arevalo:2019wdi, Crisler:2018gci,Abramoff:2019dfb,Tiffenberg:2017aac,Sensei2020} and neutrino experiments \cite{CONNIE_2019,CONNIE_2020}, a backside ohmic contact is required in order to apply the needed substrate bias to fully deplete sensor \cite{Holland:2003}. At the same time, different processing techniques are used on the backside to reduce dark current. The backside processing of these sensors determines the field shaping near the surface, and has a large impact in the CCE for events in that region. We study here the CCE for back-illuminated detectors with more than 200 $\mu$m thickness.

\section{Determination of the backside CCE using X-rays}

\begin{figure}[htbp]
\includegraphics[width=1.\textwidth]{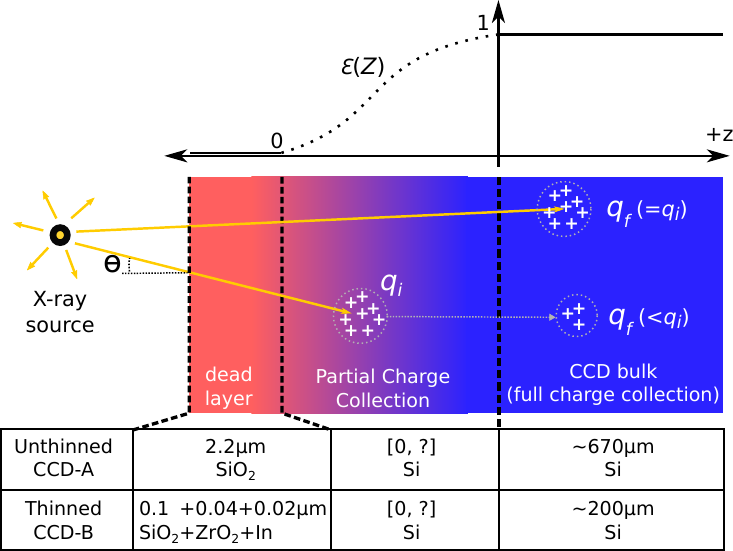}
\caption{Sketch of the CCD back illumination with an X-ray source. The photon penetrates into the CCD  producing a cloud of charge $q_i$, some fraction $\varepsilon(Z)$ of this charge gets collected depending on the depth $Z$. The region near the back of the CCD where  0<$\varepsilon(Z)$<1 is the PCC layer. }
\label{fig:cartoon}
\end{figure}

X-rays can be used to characterize the CCE near the back surface of a CCD. Figure \ref{fig:cartoon} shows a cartoon of X-ray setup together with the most important variables that participate in the analysis. Some important aspects and definitions
\begin{itemize}

\item The source emits photons with uniform angular distribution covering a full hemisphere. The angular distribution on the sensor depends on the geometry of the setup. We model the angular distribution by $f_{\Theta}(\theta)$, where $\Theta$ is measured as the angle of the incidence of the photon in the CCD compared to perpendicular direction to the back surface of the sensor.

\item The X-ray photons can reach the PCC layer and the bulk of the sensor volume. The interaction depth $Z$ in the sensor depends on the incident angle and its probability distribution function (\textit{pdf}) can be written as $f_Z(z|\theta)=(\mathrm{cos}(\theta)/\lambda)\mathrm{exp}(-z\mathrm{cos}(\theta)/\lambda)$, where $\lambda$ is the attenuation length of the photon.

\item X-rays produce an ionization charge packet with mean value $q_i = E_i/\epsilon$, where $E_i$ is the energy of the photon and $\epsilon$= 3.75 eV is the mean ionizing energy \cite{Rodrigues2020}. For now, we assume that the initial charge packet is the same for all photoelectric absorption events, we discuss later how the Fano noise affects the final results. The primary charge ionization is the same for the PCC layer and the bulk of the sensor as represented in Fig. \ref{fig:cartoon}.

\item $\varepsilon(z)$ is the CCE function in the backside of the detector. The function indicates the fraction of carriers that are collected by the pixel after drifting away from the PCC layer (carriers that do not recombine in the PCC layer). This function depends on the depth of the interaction. If the primary charge packet occurs deep in the PCC layer (far from the bulk of the sensor), carrier will have more time to recombine before they reach the bulk. Thus,  $\varepsilon(z)$ increases monotonically.

\item $q_f$ is the charge that escapes from the PCC layer and can be collected and measured by the sensor. As illustrated in Fig. \ref{fig:cartoon}, this  will depend on the interaction depth of the photon. We will refer as $Q_f$ to the random variable accounting for the possible values of the X-ray events with \textit{pdf}  $f_{Q_f}(q_f)$. The distribution of $Q_f$ is the observable in our data.

\end{itemize}    

From the previous definitions the measured charge can be expressed as 
\begin{equation}
Q_f = q_i\varepsilon(Z).    
\label{eq: Qf definition}
\end{equation}

\subsection{Determination of efficiency function using monochromatic X-ray source}
\label{sec: method}
The measured spectrum of events normalized by the total number of events ($N_T$) is an estimation $\hat{f}_{Q_f}(q_f)$ of $f_{Q_f}(q_f)$. We can then use it to estimate the cumulative distribution function (\textit{cdf}) of $Q_f$:

\begin{equation}
\label{eq: equal cum probs}
\begin{split}
    P(Q_f\le q_f) \approx 
    \hat{F}_{Q_f}(q_f) &= \int_{0}^{q_f}\hat{f}_{Q_f}(x)dx .
    \end{split}
\end{equation}
Using Eq.\eqref{eq: Qf definition} and due to the monotonically increasing  $\varepsilon(Z)$,
\begin{equation}
\label{eq: equal cum probs b}
\begin{split}
    P(Q_f\le q_f) = P(q_i\varepsilon(Z)\le q_f)=F_{Z}(z_0)= \int_{0}^{z_0}\hat{f}_{Z}(z)dz 
    \end{split}
\end{equation}


 
\noindent where $z_0$ is such that $\varepsilon(z_0)=q_f/q_i$, and
 
\begin{equation}
\label{eq: cumulative equality}
    \hat{F}_{Q_f}(q_f)=F_{Z}(z_0).
\end{equation}

The measurements at low charge values are often affected by readout noise. In this case, we calculate the \textit{cdf} integrating away from low charge values, 

\begin{equation}
    \hat{F}^{\leftarrow}_{Q_f}(q_f) = \int_{q_f}^{q_i}\hat{f}_{Q_f}(x)dx \mathrm{,\ and\ } F^{\leftarrow}_{Z}(z_0) =  \int_{z_0}^{\infty}\hat{f}_{Z}(z)dz.
    \label{eq: cumulative equality from right}
\end{equation}
For each $q_f$, we find $z_0$ such that $\hat{F}^{\leftarrow}_{Q_f}(q_f) = F^{\leftarrow}_{Z}(z_0)$ where the efficiency is $\varepsilon(z_0)=q_f/q_i$.

The method to calculate the CCE using one X-ray peak is summarized in the Table \ref{tab: one peak method} of the Appendix.

\subsection{Determination of efficiency function using an $^{55}$Fe source}\label{sec:method}

$^{55}$Fe X-ray source has an extensive use in the calibration of typical performance parameters of CCDs and other sensors \cite{janesick2001scientific}. In this article we extend its use to  characterize the charge collection in the PCC layer using the methodology proposed in Section \ref{sec: method}. The main characteristics of the three X-rays emitted by $^{55}$Fe are summarized in Table \ref{tab:table1}. $K_{\alpha}$ X-rays have similar energy and attenuation length and therefore can be treated as a single X-ray line for the purpose of this analysis.

\begin{table}
\caption{\label{tab:table1} $^{55}$Fe X-rays energies, Intensity in photons per 100 disintegrations and attenuation length in $\mu$m \cite{TabRad_v3}. Mean e-h pairs production using the mean ionization energy}.
\begin{tabular}{ccccc}
\hline
X$_K$ & Energy (keV)  & Mean e-h production ($q_i$) & Intensity  & Attenuation length ($\lambda_{\alpha}$)    \\
\hline
$\alpha_2$      & 5887.65 & 1570 & 8.45  (14)  & 28.7   \\ 
$\alpha_1$      & 5898.75 & 1573 & 16.57 (27)  & 28.9   \\
$\beta_{3}$     & 6490.45 & 1731 & 3.40  (7)   & 38.0   \\
\hline
\end{tabular}
\end{table}

Then,  \textit{pdf} for the interaction as a function of depth are 
\[
f_{Z_{\alpha}}(z|\theta)=(\mathrm{cos}(\theta)/\lambda_{\alpha})exp(-z\mathrm{cos}(\theta)/\lambda_{\alpha})f_{\Theta}(\theta)
\]
and 
\[
f_{Z_{\beta}}(z|\theta)=(\mathrm{cos}(\theta)/\lambda_{\beta})exp(-z\mathrm{cos}(\theta)/\lambda_{\beta})f_{\Theta}(\theta)
\]
for the $X_{K_{\alpha}}$ and $X_{K_{\beta}}$, respectively. With the same angular distribution in both cases. 


Generalizing Eq. \eqref{eq: equal cum probs} and \eqref{eq: cumulative equality} for two X-ray energies $\alpha$ and $\beta$, the measured \textit{cdf} for $Q_f$ is
\begin{equation}
\label{eq: two peaks cum probs}
    \hat{F}_{Q_f}(q_f) =P(Q_f\le q_f)=p_{\alpha}P(Z_{\alpha}\le z_{\alpha,0}) +p_{\beta}P(Z_{\beta}\le z_{\beta,0}),
\end{equation}

\noindent where $\varepsilon(z_{\alpha,0}) =q_f/q_{i,\alpha}$ and $\varepsilon(z_{\beta,0}) = q_f/q_{i,\beta}$, such that the depth \textit{cdf} equals the measured cumulative distribution of events. $p_{\alpha}$ and $p_{\beta}$ are the relative intensities determined by Table \ref{tab:table1} normalized by the number of desintegrations. Since we assume a monotonically increasing $\varepsilon(z)$ function, then $z_{\alpha,0} \ge z_{\beta,0}$. Using a more condense notation  
\begin{equation}
\label{eq: two peaks cumulative equality}
    \hat{F}_{Q_f}(q_f)=p_{\alpha}F_{Z_{\alpha}}(z_{\alpha,0})+p_{\beta}F_{Z_{\beta}}(z_{\beta,0})  \mathrm{\, with\ } z_{\alpha,0} \ge z_{\beta,0}.
\end{equation}

A recursive nonlinear numeric solver is used to find $z_{\alpha,0}$ and $z_{\beta,0} $ simultaneously. Three features of the $^{55}$Fe source can be used to simplify the problem. 
\begin{itemize}
    \item Larger $X_{K_{\alpha}}$-flux than $X_{K_{\beta}}$-flux, since $p_{\alpha}/p_{\beta} = 7.47$
    \item $F_{Z_{\alpha}}(z_{\alpha,0})$ is always greater than $F_{Z_{\beta}}(z_{\beta,0})$ because of the difference in the attenuation length ($\lambda_{\alpha}<\lambda_{\beta}$) and the fact that $z_{\alpha,0} \ge z_{\beta,0}$.  
    \item As $q_f$ becomes smaller than $q_{i,\alpha}$ then $q_f/q_{i,\alpha}$ becomes closer to $q_f/q_{i,\beta}$, and therefore $z_{\alpha,0}$ becomes closer to $z_{\beta,0}$. In fact, $q_{i,\beta}$ and $q_{i,\alpha}$ differ only 10\%. 
\end{itemize}

Most of the signal is dominated by the $X_{K_{\alpha}}$ photon and a small effect is introduced by assuming a unique $z_0= z_{\alpha,0}=z_{\beta,0}$. This assumption allows to follow the same procedure presented in Section \ref{sec: method} to solve equation \ref{eq: two peaks cumulative equality}.
Assuming $z_0= z_{\alpha,0}=z_{\beta,0}$ the true collection efficiency at $z_0$ lays  between $q_f/q_{i,\beta}$ and $q_f/q_{i,\alpha}$. A simple approximation is $\varepsilon(z_0)=q_f/(p_{\alpha}q_{i,\alpha} + p_{\beta}q_{i,\beta})$. The full method for an $^{55}$Fe source is summarized in Table \ref{tab: two peaks method}, in the Appendix.

\section{Experimental results}

We study here two different CCDs. 

CCD-A  was designed by the LBNL Microsystems Laboratory \cite{LBNL-MSL} as part of the R\&D effort for low energy neutrino experiments \cite{CONNIE_2019} and low mass direct dark matter search \cite{Aguilar-Arevalo:2019wdi}. This is a rectangular CCD with 8 million square pixels of 15 $\mu$m $\times$ 15 $\mu$m each. The CCD  is fabricated in n-type substrate with a full thickness of 675 $\mu$m. The  resistivity of the substrate  greater than 10000 $\Omega$-cm. The CCD is operated with 40V bias voltage that fully depletes the high-resistivity substrate using the method developed in Ref.\cite{Holland:2003}.  In order to trap impurities that migrate during the sensor processing, a 1$\mu$m thick  in-situ doped polysilicon (ISDP) layer is deposited on the backside of the detector. This layer plays a critical role controlling the dark current of the detector. Additional layers of silicon nitride, phosphorous-doped polysilicon and silicon dioxide are added to the backside (~2 $\mu$m total thickness). Phosphorous can migrate into the high resistivity material producing a region of a few microns where charge can recombine before drifting to the collecting gates of the detector. This region constitutes the PCC layer that we characterize with $^{55}$Fe X-rays, as shown in Figure \ref{fig:cartoon}.

CCD-B is similar to CCD-A with a few important differences. The detector has 4 million pixels, with a thickness of 200 $\mu$m. It is also fabricated in high resistivity n-type silicon. The backside of the sensor has been processed for astronomical imaging. A backside ohmic contact is formed by low-pressure, chemical-vapor deposition in-situ doped polycrystalline silicon (ISDP). 
This layer is  made thin for good blue response, typically 10-20 nm, and is robust to over-depleted operation that is necessary to guarantee full depletion across the entire CCD. This detector is operated at bias voltage of 40 V. Because of its backside treatment, this detector is not expected to have significant charge recombination near the back surface. The detector is exposed to $^{55}$Fe X-rays on the backside, as shown in Figure \ref{fig:cartoon}.

The $^{55}$Fe was located 3.55 cm away from the CCDs. The effective depth distribution of interacting photons was calculated using a Monte Carlo simulation, and the result is 

\begin{equation}
    \label{eq: depth distribution thick detectors}
    f_{Z}(z) = I_\alpha*exp(-z/\tau_\alpha)+I_\beta*exp(-z/\tau_\beta),
\end{equation}
where $I_\alpha$($I_\beta$) represents the intensity and $\tau_\alpha$($\tau_\beta$) is the effective optical depth for the $\alpha$($\beta$) spectral line. $I_\alpha = 0.034$, $I_\beta = 0.0032$, $\tau_\alpha = 25.74 \mu$m, and $\tau_\beta = 37.19 \mu$m.

\subsection{Results for CCD-A}

\begin{figure}[htbp]
\includegraphics[width=0.49\textwidth]{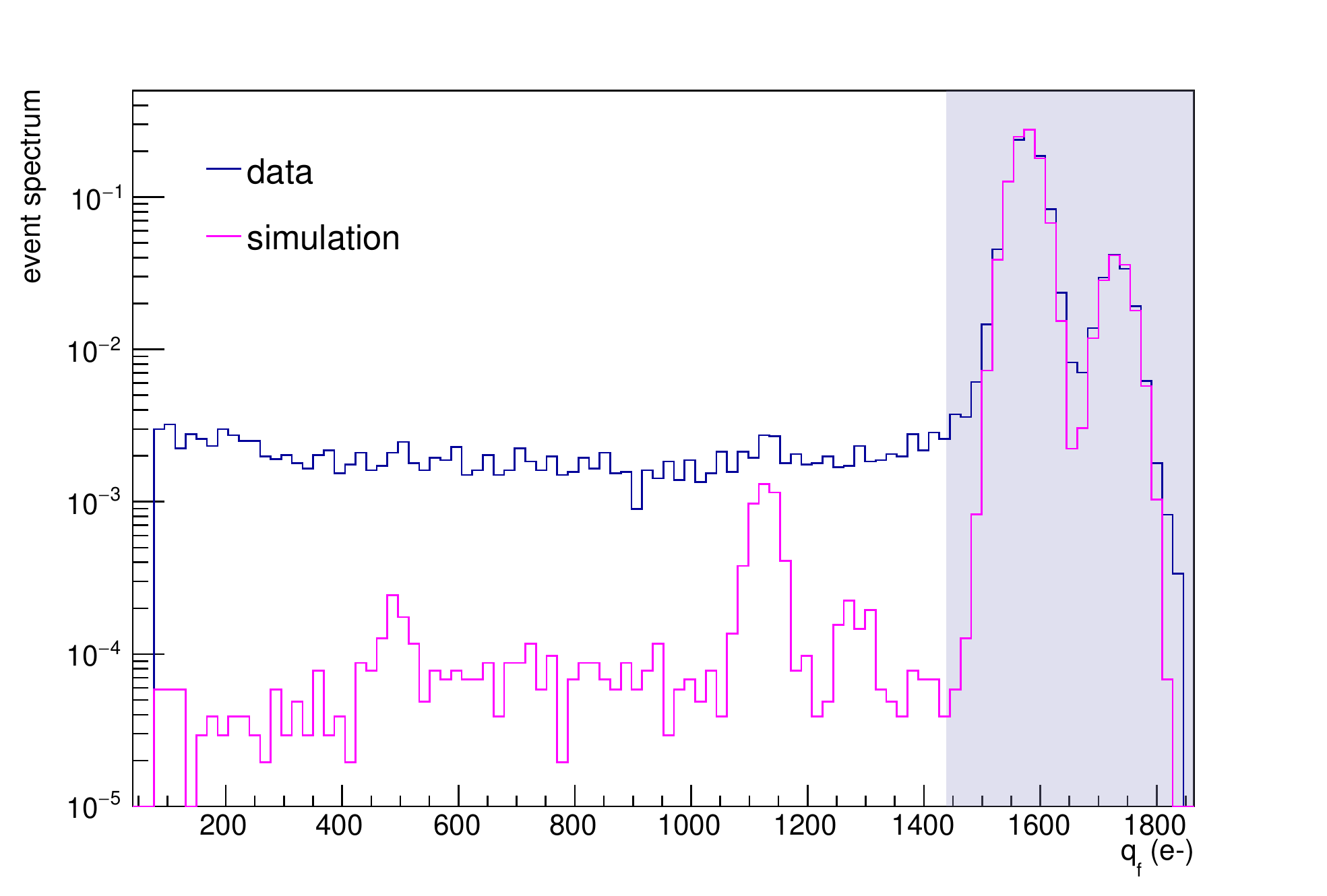}
\includegraphics[width=0.49\textwidth]{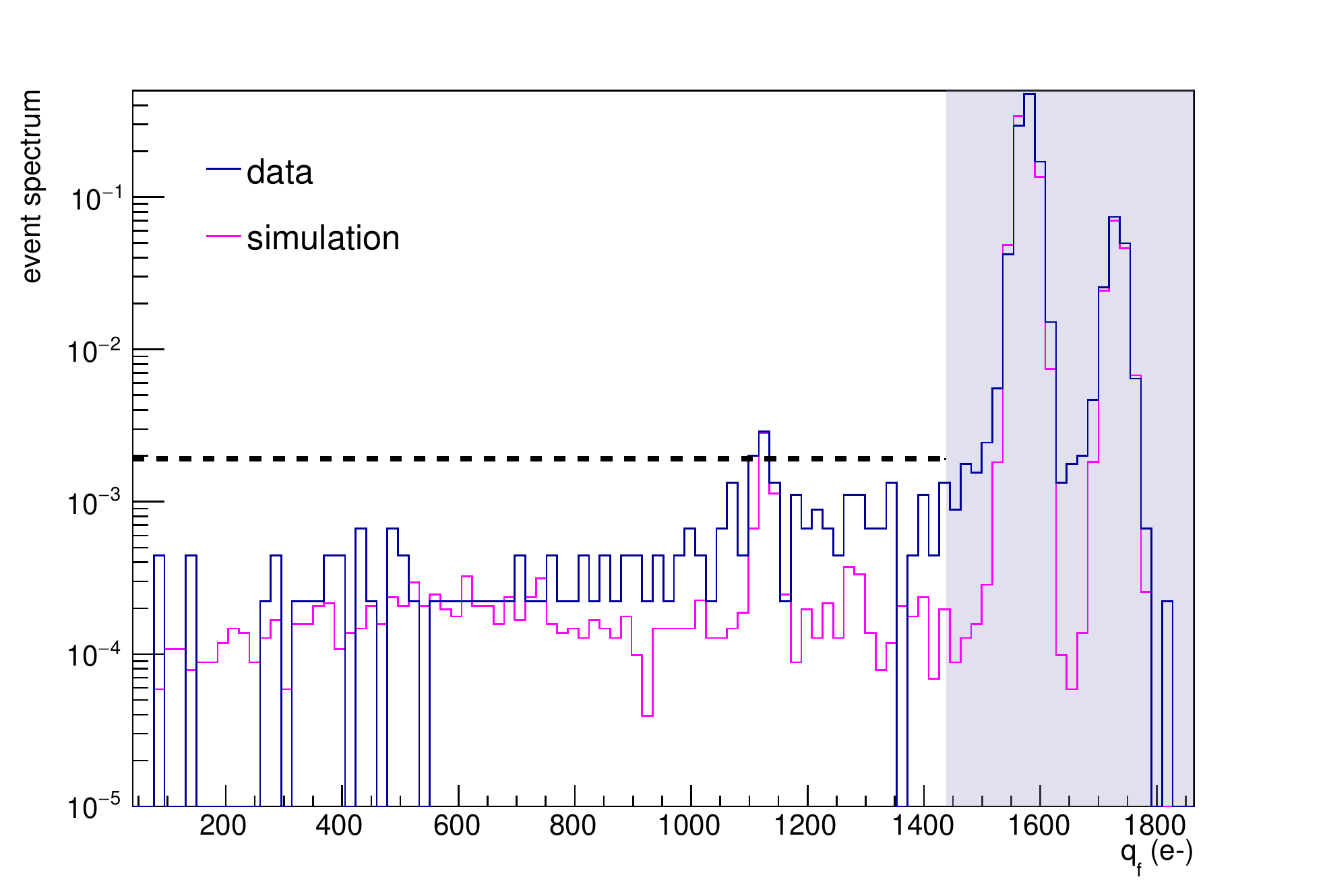}
\caption{Event spectra for CCD-A (left) and CCD-B (right) calculated using bin size of 70 eV normalized by the number of measured events in the $K_{\alpha}$ peak. Blue: measured spectra. Magenta: Simulated spectra of events from Geant4. Left figure: spectra for CCD-A; 35195 events in the histogram; 26697 events in the $K_{\alpha}$ peak. Right figure: event spectra for CCD-B; 5452 events in the histogram; 4482 events in the $K_{\alpha}$ peak. The dashed black line indicates the expected level of events if the partial charge collection layer on CCD-B was same as the one measured on CCD-A.}
\label{fig:measured spectrum thick ccd}
\end{figure}

\begin{figure}[htbp]
\includegraphics[width=1.\textwidth]{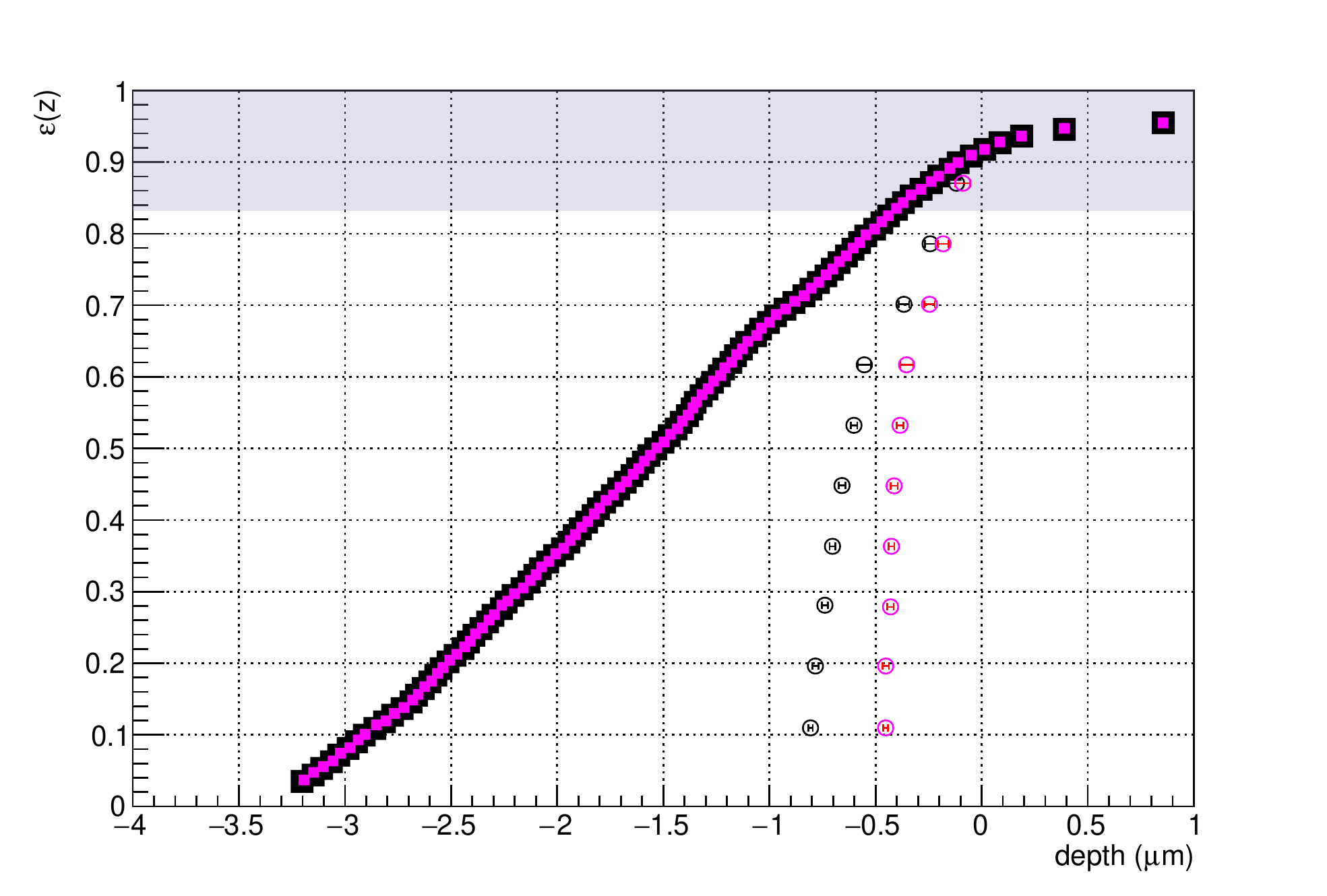}
\caption{Measured charge collection efficiency for a CCD-A (solid square markers) and CCD-B (open circle markers). The black points show the results without considering the background events predicted by the simulation. The magenta point shows the results after correcting the experimental spectra by subtracting the events form the simulations. The shaded area indicates the region where the detailed shape of the X-ray peaks affect the measurement, introducing more uncertainty.}
\label{fig:measured efficiency thick ccd}
\end{figure}

The spectrum of measured charge for CCD-A is shown in the top panel of Fig. \ref{fig:measured spectrum thick ccd}, and compared with a Geant4\cite{GEANT4} simulation assuming perfect CCE for the entire volume of the sensor ($\varepsilon(z)=1$). The K$_\alpha$ and K$_\beta$ peaks from Table \ref{tab:table1} are evident. The excess of reconstructed events to the left of these peaks is attributed to the PCC layer, where charge recombination produces a measurement below the peak energy. The bump in the simulation around 1100 e$^-$ is an escape peak, as discussed in Ref.\cite{Jaeckel:2010xx}. This data is used to measure the CCE function $\varepsilon(z)$ following the prescription in Section \ref{sec:method}, and the results are shown in the top panel of Fig.\ref{fig:measured efficiency thick ccd}. The depth scale is chosen such that $ \varepsilon(z=0) = 0.9$.
The shaded region corresponds to the energies between 5.4~keV and 7~keV where the events from $K_{\alpha}$ and $K_{\beta}$ are dominant and systematic uncertainties are expected to be important. In this region the precise shape of $\varepsilon$ curve is less reliable.

\subsection{Results for CCD-B}

The spectrum of measured charge for CCD-B is shown in the bottom panel of Fig.\ref{fig:measured spectrum thick ccd}, and compared with a Geant4 \cite{GEANT4} simulation with perfect CCE. As for CCD-A, the K$_\alpha$ and K$_\beta$ spectral lines are evident,  CCD-B has a different output stage producing higher resolution peaks \cite{Tiffenberg:2017aac}.  The relative rate of events on the left of the peaks, are well below the rate observed for CCD-A and consistent with the simulation. These events are produced mostly by low probability Compton scattering of X-rays. The CCE function $\varepsilon(z_{\alpha})$ is determined as discussed in Section \ref{sec:method} and the results are shown in Fig.\ref{fig:measured efficiency thick ccd}  bottom panel, black circles. The measurement of $\varepsilon(z_{\alpha})$ is also performed after the predicted Compton spectrum is subtracted based on the simulation, the results are shown in  bottom panel of Fig.\ref{fig:measured efficiency thick ccd}, magenta circles. As before, the horizontal axis is selected such that $ \varepsilon(z=0) = 0.9 $.

\subsection{Conclusion}

The results of CCD-A and CCD-B showed in Fig.\ref{fig:measured efficiency thick ccd} demonstrate the large impact that the backside processing could have in the CCE for back-illuminated detectors. When a layer of a few microns with charge recombination is present on the CCD, the spectrum for low energy X-rays gets significantly distorted. The charge recombination generates a significant number of lower energy events in the spectrum. The backside processing performed in detectors optimized for astronomical instruments eliminates this issue for the most part, as shown with CCD-B. The generation of low energy events constitute a major concern for experiments looking for rare signals near the detector threshold~\cite{Aguilar-Arevalo:2016ndq, Aguilar-Arevalo:2016zop,Aguilar-Arevalo:2019wdi, Crisler:2018gci,Abramoff:2019dfb,Tiffenberg:2017aac,Sensei2020,CONNIE_2019,CONNIE_2020}.

The results obtained here for CCD-B, optimized for astronomical imaging, are consistent with the observations of detection efficiency and reflectivity in Ref.\cite{LBNLQE}. 

A new technique was introduced here to characterize the CCE for back-illuminated CCDs, this technique can easily be generalized to other semiconductor detectors. The technique uses tools that are commonly available at the detector characterization laboratories. As shown here, the new method is capable of measuring a PCC layer of a few micrometers. The sensitivity to a very thin PCC layer is limited by the energy of the $^{55}$Fe X-rays, and the technique could be easily extended for much thinner recombination layers using lower energy X-rays. This technique will be a powerful tool in the optimization of detectors for the next generation of low threshold experiments looking for rare events such as dark matter, or coherent neutrino nucleus scattering\cite{Oscura,Violeta}.

\appendix
\section*{Appendix: Details of the method}

The details of the method to measure the CCE in the backside of a back-illuminated CCD are presented in Table \ref{tab: one peak method}. The details of method used with the $^{55}$Fe source having two X-ray lines is presented in Table \ref{tab: two peaks method}.

\begin{table}
\begin{center}
\begin{tabularx}{\textwidth}[t]{X}
\noalign{\hrule height 2pt}

\textbf{1) Calculate angular distribution of incident photons:} \\
\hline
Based on the geometry of the experiment evaluate $f_{\Theta}(\theta)$.\\
\hline
\textbf{2) Calculate depth distribution of events:} \\
\hline
$f_Z(z|\theta)=(\mathrm{cos}(\theta)/\lambda) exp(-z\mathrm{cos}(\theta)/\lambda)f_{\Theta}(\theta)$, where $\lambda$ is the attenuation length of the photon. Then, calculate the \textit{cdf} $F_{Z}(z_0)$ (or $F^{\leftarrow}_{Z}(z_0)$ from Eq. \eqref{eq: cumulative equality from right}).
\\
\hline
\textbf{3) Make a spectrum of measured events:} \\
\hline
Calculate the spectrum of events reconstructed from the data and normalize it by total number of events ($N_T$). This is the estimation $\hat{f}_{Q_f}(q_f)$.\\
\hline
\textbf{4) Calculate integral of the measured spectrum up to a charge $q_f$:}  \\
\hline
Calculate \textit{cdf} either $\hat{F}_{Q_f}(q_f)$ (from Eq. \ref{eq: equal cum probs}), or $\hat{F}^{\leftarrow}_{Q_f}(q_f)$ (from Eq. \ref{eq: cumulative equality from right}).
\\
\hline
\textbf{5) Find $z_0$:}  \\
\hline
Find $z_0$ that equals the \textit{cdf} of the interaction depth with the cumulative proportion of measured events. This is $\hat{F}_{Q_f}(q_f)=F_{Z}(z_0)$, or $\hat{F}^{\leftarrow}_{Q_f}(q_f) = F^{\leftarrow}_{Z}(z_0)$.
\\
\hline
\textbf{6) Calculate the efficiency at $z_0$:}  \\
\hline
$\varepsilon(z_0)=q_f/q_i$.
\\
\hline
\textbf{7) Repeat steps 4, 5 and 6 for a different $q_f$ to complete $\varepsilon(z)$.}\\
\noalign{\hrule height 2pt}

\end{tabularx}

\end{center}
\caption{Methodology to calculate the PCC efficiency function using one X-ray peak.}
\label{tab: one peak method}
\end{table}

\begin{table}
\begin{center}
\begin{tabularx}{\textwidth}[t]{X}
\noalign{\hrule height 2pt}

\textbf{1) Calculate angular distribution of incident photons:} \\
\hline
Based on the geometry of the experiment evaluate $f_{\Theta}(\theta)$.\\
\hline
\textbf{2) Calculate depth distribution of events:} \\
\hline
$f_Z(z|\theta)=(p_{\alpha}(\mathrm{cos}(\theta)/\lambda_{\alpha})exp(-z\mathrm{cos}(\theta)/\lambda_{\alpha}) + p_{\beta}(\mathrm{cos}(\theta)/\lambda_{\beta})exp(-z\mathrm{cos}(\theta)/\lambda_{\beta}))f_{\Theta}(\theta)$, where $\lambda$ is the attenuation length of the photon. Then, calculate the cumulative distribution $F_{Z}(z_0)$ (or $F^{\leftarrow}_{Z}(z_0)$ from equation \ref{eq: cumulative equality from right}).
\\
\hline
\textbf{3) Make an spectrum of measured events:} \\
\hline
Calculate the spectrum of events reconstructed from the data and normalize it by total number of events ($N_T$). This is the estimation $\hat{f}_{Q_f}(q_f)$.\\
\hline
\textbf{4) Calculate integral of the measured spectrum up to a charge $q_f$:}  \\
\hline
Calculate cumulative distributions either $\hat{F}_{Q_f}(q_f)$ (from Eq. \ref{eq: equal cum probs}), or $\hat{F}^{\leftarrow}_{Q_f}(q_f)$ (from Eq. \ref{eq: cumulative equality from right}).
\\
\hline
\textbf{5) Find $z_0$:}  \\
\hline
Find $z_0$ that equals the \textit{cdf} of the interaction depth with the cumulative proportion of measured events. This is $\hat{F}_{Q_f}(q_f)=F_{Z}(z_0)$, or $\hat{F}^{\leftarrow}_{Q_f}(q_f) = F^{\leftarrow}_{Z}(z_0)$.
\\
\hline
\textbf{6) Calculate the efficiency at $z_0$:}  \\
\hline
$\varepsilon(z_0)=q_f/(p_{\alpha}q_{i,\alpha} + p_{\beta}q_{i,\beta})$
\\
\hline
\textbf{7) Repeat steps 4, 5 and 6 for a different $q_f$ to complete $\varepsilon(z)$.}\\
\noalign{\hrule height 2pt}

\end{tabularx}

\end{center}
\caption{Methodology to calculate the partial charge collection efficiency function using $^{55}$Fe source.}
\label{tab: two peaks method}
\end{table}

\acknowledgments

We thank the SiDet team at Fermilab for the support on the operations of CCDs and Skipper-CCDs, specially Kevin Kuk and Andrew Lathrop. We are grateful to Oscar von Uri for taking care of no-solo-bar problem.
This work was supported by Fermilab under DOE Contract No.\ DE-AC02-07CH11359. 
This manuscript has been authored by Fermi Research Alliance, LLC under Contract No. DE-AC02-07CH11359 with the U.S.~Department of Energy, Office of Science, Office of High Energy Physics. The United States Government retains and the publisher, by accepting the article for publication, acknowledges that the United States Government retains a non-exclusive, paid-up, irrevocable, world-wide license to publish or reproduce the published form of this manuscript, or allow others to do so, for United States Government purposes.




\bibliographystyle{apsrev4-1}
\bibliography{biblio.bib}

\begin{thebibliography}{25}
\bibitem [{Osc()}]{Oscura}%
  \BibitemOpen
  \href@noop {} {}\bibinfo {note} {The Oscura project is an R\&D effort
  supported by Department of Energy to develop 10 kg skipper-CCD dark matter
  search.}\BibitemShut {Stop}%
\end{thebibliography}



\end{document}